\documentclass[aps,showpacs,11pt]{revtex4}
\usepackage{dcolumn}
\usepackage{graphicx}
\usepackage{amsmath}
\usepackage{amsfonts}
\usepackage{amssymb}
\usepackage{psfrag}
\usepackage{wrapfig}
\usepackage{subfigure}
\usepackage{makeidx}
\usepackage{bm}
\usepackage{epsf}
\usepackage{hyperref}
\usepackage{color}

\begin{document}

\title{Surface term, Virasoro algebra, and Wald entropy of black holes in higher-curvature gravity}

\author{Shao-Jun Zhang}
\email{sjzhang84@sjtu.edu.cn}
\author{Bin Wang}
\email{wang$_$b@sjtu.edu.cn}
\affiliation{Department of
Physics, INPAC, Shanghai Jiao Tong University, Shanghai 200240, China}
\date{\today}

\begin{abstract}
\indent Recently, in the Einstein gravity, Majhi
and Padmanabhan proposed a straightforward and
transparent way of obtaining the
Bekenstein-Hawking entropy by using an approach
based on the Virasoro algebra and central charge.
In this work, we generalize their approach to the
modified gravity with higher-curvature
corrections and show that their approach can
successfully lead to the corresponding Wald
entropy in the higher-curvature gravity. Our
result shows that the approach is physically
general.

\end{abstract}


\maketitle
\section{Introduction}

It is well known that black holes act as a bridge connecting general
relativity and quantum mechanics. Studying  various properties of
black holes can provide us with a possible way to understand the
underlying quantum gravity. One of the most remarkable features of a
black hole is its thermodynamical ensemble
~\cite{Bekenstein:1973a,Bardeen:1973a,Hawking:1974a,Hawking:1975a},
which has the concept of  not only temperature, but also entropy.  In
classical general relativity, it is usually hard to have a statistical understanding of black hole entropy. In the past
decades, various attempts have been made to interpret the
microscopic origin of black hole entropy. One of them,
is to use Virasoro algebra and the Cardy
formula~\cite{Cardy:1986a,Bloete:1986a,Carlip:1998a} to explain the
statistical meaning of the entropy~\cite{Carlip:1998b,Carlip:1999a}.
This method was based on the original approach by Brown and
Henneaux~\cite{Brown:1986a} in studying the asymptotical symmetry
group of ($2+1$)-dimensional gravity in asymptotically anti-de Sitter
spacetime. The basic idea of this method is to define the
corresponding Noether charges and Lie bracket among  some chosen
diffeomorphisms; then the Lie bracket turns out to be the
well-known Virasoro algebra with central extension. One can read off
the deduced central charge and the zero-mode eigenvalues of the
Fourier modes of the charge. Substituting them into the Cardy
formula, Bekenstein-Hawking entropy can be obtained. Recent
development and related references can be found in
Ref.~\cite{Majhi:2011a} and references therein. In this approach,
the Noether current was constructed starting from the bulk action,
the Einstein-Hilbert action, or the Lovelock action.

Recently, a new approach was proposed by Majhi
and Padmanabhan \cite{Majhi:2012a}
\cite{Majhi:2012b}.  They introduced the Virasoro
algebra defined by the Noether currents
corresponding to the surface term of
gravitational action, the Gibbons-Hawking
term~\cite{Gibbon:1977a},  instead of the bulk
action. In their work, the diffeomorphisms
related to the Noether currents are chosen to be
those which leave the near-horizon form of the
metric invariant in some nonsingular
coordinates. This new approach is clear and
straightforward for obtaining the correct entropy,
and does not require any {\em ad hoc} prescriptions,
such as shifting the zero-mode energy~\cite{Carlip:1999a} or choosing a parameter in
the Fourier modes of $T$ as the surface gravity~\cite{Silva:2002a}, etc. In the Einstein gravity,
this approach leads to the Bekenstein-Hawking
entropy  without any ambiguities.

The Bekenstein-Hawking entropy, in proportion to
the area of the black hole horizon, holds well in
Einstein gravity; however, it does not hold in
general gravitational theory. For example, in the
gravity theory with higher-curvature corrections,
the entropy is described by Wald entropy, which
contains properly the contributions of higher-curvature
corrections~\cite{Wald:1993a,Iyer:1994a,Iyer:1995a,Jacobson:1993a}.
It is natural to ask whether the  method proposed
by Majhi and Padmanabhan can be generalized to
produce the Wald entropy in the gravity theory
with higher-curvature corrections. This will
serve as the main motivation of the present
paper. We will study the higher-curvature
gravitational theories, including
Gauss-Bonnet and third-order Lovelock gravity. We
will show that the Virasoro algebra, with the
information of higher-curvature terms defined by
the Noether currents corresponding to the surface
term of gravitational action has a central
extension that indeed leads to the Wald entropy.
We will argue that the method proposed by Majhi
and Padmanabhan is general, which can provide
deeper insight into the nature of the entropy.

The organization of the paper is as follows: In
the next section, we will give a brief review on
general formulas of Noether current constructed
from the surface term of gravitational action. In
Sec. III, we will first give general
derivations of related vector fields for static
spherical spacetimes, and then apply three
different gravitational theories, namely the
Einstein, Gauss-Bonnet, and third-order Lovelock
theories.  We will derive the charge
and Virasoro algebra starting from the
corresponding surface term of the gravitational
action and then use the Cardy formula to obtain
the entropy in different gravitational theories.
We will summarize our results in the last section.

\section{General formulas of Noether current from the surface term}

In this section, let us briefly review the
general formulas of Noether current constructed
from the surface term of gravitational
action~\cite{Majhi:2011a,Majhi:2012a,Majhi:2012b}.
Let us consider a general surface term as
\begin{eqnarray}
I_B &=& \frac{1}{8\pi G} \int_{\partial {\cal M}} d^{n-1} x \sqrt{\gamma} {\cal L}_{B}\nonumber\\
&=&\frac{1}{8\pi G} \int_{{\cal M}} d^n x \sqrt{g} \nabla_a({\cal L}_{B} N^a),\label{e20}
\end{eqnarray}
where $N^a$ is the unit normal vector of the
boundary $\partial {\cal M}$, $g_{ab}$ denotes
the bulk metric and $\gamma_{ab}$ denotes the
induced boundary metric. The Noether current we
will study corresponds to diffeomorphism
transformation. For convenience, we define $A^a
\equiv \frac{1}{8\pi G} {\cal L}_{B} N^a$, so the $n$-dimensional Lagrangian density is
$\sqrt{g} {\cal L} = \sqrt{g} \nabla_a A^a$.
Under a general diffeomorphism $x^a \rightarrow
x^a +\xi^a$, the Lagrangian density $\sqrt{g}
{\cal L}$ changes by
\begin{eqnarray}
\delta_\xi (\sqrt{g} {\cal L}) \equiv \pounds_\xi (\sqrt{g} {\cal L})=\sqrt{g} \nabla_a ({\cal L} \xi^a).\label{e21}
\end{eqnarray}
In deriving the above equation,  we have used
the formulas $\pounds_\xi (\sqrt{g}) = \sqrt{g}
\nabla_a \xi^a$ and $\pounds_\xi ({\cal L}) =
\xi^a \nabla_a {\cal L}$. On the other hand, we
also have
\begin{eqnarray}
\delta_\xi \left(\sqrt{g} \nabla_a A^a\right)&=& \pounds_\xi [\partial_a \left(\sqrt{g} A^a\right)]\nonumber\\
&=&\sqrt{g}\nabla_a\left[\nabla_b\left(A^a
\xi^b\right)-A^b \nabla_b
\xi^a\right],\label{e22}
\end{eqnarray}
where we have used the formula $\nabla_a A^a =
\frac{1}{\sqrt{g}}
\partial_a \left(\sqrt{g} A^a\right)$.

Equating Eqs.~(\ref{e21}) and~(\ref{e22}), we get
the conserved Noether current
\begin{eqnarray}
J^a [\xi]={\cal L} \xi^a -\nabla_b\left(A^a \xi^b\right) + A^b \nabla_b \xi^a.\label{e23}
\end{eqnarray}
Substituting ${\cal L} =\nabla_a A^a$ into the above equation, the Noether current can be rewritten as
\begin{eqnarray}
J^a [\xi] = \nabla_b J^{ab} [\xi] = \nabla_b \left[\xi^a A^b-\xi^b A^a\right].\label{e24}
\end{eqnarray}
Usually, $J^{ab}$ is called the Noether
potential. With the explicit expression
$A^a=\frac{1}{8\pi G}{\cal L_B} N^a$, we get the
final form of the Noether current constructed
from the surface term
\begin{eqnarray}
J^a [\xi] = \nabla_b J^{ab} [\xi] = \frac{1}{8\pi G}\nabla_b \left[{\cal L}_B \left(\xi^a N^b-\xi^b N^a\right)\right].\label{e25}
\end{eqnarray}
The corresponding charge is defined as
\begin{eqnarray}
Q [\xi] = \frac{1}{2} \int_{\Sigma} \sqrt{h} d \Sigma_{ab} J^{ab}, \label{e26}
\end{eqnarray}
where $d \Sigma_{ab} =- d^{n-1}x (N_a M_b - N_b
M_a)$ is the surface element of the
$(n-2)$-dimensional surface $\Sigma$, and $h$ is the
determinant of the corresponding induced metric.
$N^a$ and $M^a$ are chosen to be spacelike and
timelike, respectively. In our discussion, $\Sigma$
will be near the horizon ${\cal H}$.

Following Ref.~\cite{Majhi:2011a}, we define a Lie
bracket for the charges
\begin{eqnarray}
[Q_1, Q_2] :&=& \left(\delta_{\xi_1} Q[\xi_2]-\delta_{\xi_2} Q[\xi_1]\right)\nonumber\\
&=& \int_\Sigma \sqrt{h} d \Sigma_{ab} \left[\xi_2^a J^b[\xi_1]-\xi_1^a J^b[\xi_2]\right],\label{e27}
\end{eqnarray}
which will lead to the usual Virasoro algebra
with central extension. As we will see soon, with
the deduced central charge and the Cardy formula,
the entropy of black holes can be derived.

\section{Entropy of black holes from the Cardy formula}

To derive the Noether charge and Virasoro
algebra, we first need to identify appropriate
diffeomorphisms, the related vector fields
$\xi^a$. In Ref.~\cite{Majhi:2012a}, the authors
did this in the explicit Rindler metric. Here, we
give a more abstract derivation of the vector
fields $\xi^a$.

To be transparent,  we only consider static
spherical black holes with the general form of the
metric
\begin{eqnarray}
ds^2 = -f(r) dt^2+\frac{dr^2}{f(r)} + r^2 \Omega_{ij} (x) dx^i dx^j,\label{e30}
\end{eqnarray}
where $\Omega_{ij}(x)$ is the $(n-2)$-dimensional
space and $h_{ij}=r^2 \Omega_{ij}$. The horizon
is located at $r=r_h$, satisfying $f(r_h)=0$. The two
normal vector fields $N^a, M^a$ are
\begin{eqnarray}
N^a=\left(0,\sqrt{-f(\rho+r_h)},0,\cdots,0\right),\quad
M^a=\left(\sqrt{\frac{1}{f(\rho+r_h)}},0,\cdots,0\right).\label{e31}
\end{eqnarray}
To study the near-horizon structure, it is
convenient to define $r\equiv \rho +r_h$, and in
the near-horizon limit, $\rho \rightarrow 0$. The
metric then becomes
\begin{eqnarray}
ds^2 = -f(\rho+r_h) dt^2+\frac{d\rho^2}{f(\rho+r_h)} + (\rho+r_h)^2
\Omega_{ij} (x) dx^i dx^j.\label{e32}
\end{eqnarray}
The function $f(\rho+r_h)$ in the near-horizon limit
can be expanded as $f(\rho+r_h) = 2 \kappa
f'(r_h) \rho + \frac{1}{2} f''(r_h)
\rho^2+\cdots$ with surface gravity
$\kappa=\frac{f'(r_h)}{2}$. Then it is easy to
see that the $t-\rho$ part of the above metric
becomes Rindler in the near-horizon limit,
$ds^2_{t-\rho} = -2\kappa \rho dt^2 +
\frac{1}{2\kappa \rho} d\rho^2$, if we only keep
the first-order term of the expansion of
$f(\rho+r_h)$. In Ref.~\cite{Majhi:2012a}, the
authors started from a Rindler metric. Here, we
keep the full metric in the process of derivation
and only do the near-horizon limit in the end.
Introducing Bondi-like coordinates by the
transformation \cite{Majhi:2012a}
\begin{eqnarray}
du=dt-\frac{d\rho}{f(\rho+r_h)},\label{e33}
\end{eqnarray}
the metric is transformed to
\begin{eqnarray}
ds^2=-f(\rho+r_h) du^2 -2 du d\rho+ (\rho+r_h)^2 \Omega_{ij} (x)
dx^i dx^j.\label{e34}
\end{eqnarray}
To obtain the Noether current, we choose the
vector fields $\xi^a$ which can leave the horizon
structure invariant
\begin{eqnarray}
0=\pounds_\xi g_{\rho\rho} &=& -2 \partial_\rho \xi^\mu,\nonumber\\
0=\pounds_\xi g_{u\rho} &=&
-f(\rho+r_h)\partial_\rho\xi^u-\partial_\rho \xi^\rho-\partial_u
\xi^u.\label{e35}
\end{eqnarray}
Solving the above equations, we obtain
\begin{eqnarray}
\xi^u=F(u,x),\quad \xi^\rho=-\rho \partial_u F(u,x),\label{e36}
\end{eqnarray}
while the other components vanish. It is easy to
check that $\pounds_\xi g_{uu} = {\cal O}
(\rho)$, so that the condition $\pounds_\xi
g_{uu} = 0$ is  satisfied near the horizon. This
shows that the  derived vector fields are indeed
what we want. Returning to the original
coordinates $(t,\rho)$, these vector fields
become
\begin{eqnarray}
\xi^t = T-\frac{1}{f(r_h+\rho)} \partial_t T,\quad \xi^\rho = -\rho
\partial_t T,\quad T(t,\rho,x)\equiv F(u,x).\label{e37}
\end{eqnarray}
For a given $T$, from Eqs.~(\ref{e37}) and
(\ref{e26}), we have expressions of the
corresponding vector field $\xi^a$
 and the charge $Q$, respectively.

Expanding $T$ in terms of the basis functions $T_m$
\begin{eqnarray}
T=\sum_m A_m T_m,\qquad
A_m^\ast=A_{-m}\label{e38}
\end{eqnarray}
and substituting it into Eqs.~(\ref{e37})
and~(\ref{e26}), we get corresponding expansions
for $\xi^a$ and $Q$ in the forms of Eqs.~(\ref{e37})
and~(\ref{e26}), where $T$ is replaced by $T_m$.
$T_m$ is chosen to be the basis that yields the
resulting $\xi^a_m$, obeying the algebra
isomorphic to Diff $ S^1$:
\begin{eqnarray}
i\{\xi_m, \xi_n\}^a = (m-n) \xi^a_{m+n},\label{e39}
\end{eqnarray}
with $\{,\}$ being the Lie bracket. A standard choice
is~\cite{Silva:2002a}
\begin{eqnarray}
T_m=\frac{1}{\alpha} \exp\left[im(\alpha t +g(\rho) +p\cdot
x)\right],\label{e310}
\end{eqnarray}
where $p$ is an integer and $g(\rho)$ as a
function that is regular at the horizon. The value $\alpha$
is a parameter which needs to be chosen
according to the surface gravity in Refs.~\cite{Silva:2002a,Majhi:2011a}, but it is
arbitrary in the present approach and will not
affect the final results.

\subsection{Warmup: Four-dimensional Schwarzschild black hole}

As a warmup, let us first apply the above
formulas to a four-dimensional Schwarzschild black
hole in Einstein gravity. In the local Rindler
frame, the results were shown in Refs.~\cite{Majhi:2012a}\cite{Majhi:2012b}.

The surface term of the gravitational action is
the well-known Gibbons-Hawking
term~\cite{Gibbon:1977a}
\begin{eqnarray}
{\cal L}_B= K,\label{e3A0}
\end{eqnarray}
with $K=-\nabla_a N^a$ being the trace of the extrinsic
curvature of the boundary $\partial {\cal M}$.
Considering $f(r)=1-\frac{2M}{r}$ and
$\Omega_{ij}dx^i dx^j=d\Omega_2^2$ for the
Schwarzschild black hole, we have the horizon
${\cal H}$ at $r_h=2M$, and the new form of the
metric becomes
\begin{eqnarray}
ds^2 = -\frac{\rho}{2M +\rho} dt^2 + \frac{2M +\rho}{\rho} d\rho^2 + (2M+\rho)^2 d\Omega_2^2,\label{e3A2}
\end{eqnarray}
where $\rho = 0$ represents the horizon. In
Bondi-like coordinates defined by the
transformation [Eq.~(\ref{e33})], we have
\begin{eqnarray}
d u = dt - \frac{2M+\rho}{\rho} d\rho, \label{e3A3}
\end{eqnarray}
and the metric becomes
\begin{eqnarray}
ds^2=-\frac{\rho}{2M +\rho} du^2 -2 du d\rho+ (2M +\rho)^2 d\Omega_2^2.\label{e3A4}
\end{eqnarray}
In the coordinates $(t, \rho)$, $\xi^a$ are
chosen to be
\begin{eqnarray}
\xi^t = T - (2M +\rho) \partial_t T,\quad \xi^\rho=-\rho \partial_t T,\label{e3A5}
\end{eqnarray}
where $T(t,\rho,x)\equiv F(u,x)$.

Now we are ready to derive the Noether current,
and then the Virasoro algebra. For the
Schwarzschild black hole metric, we have
\begin{eqnarray}
N^a&=&\left(0,\sqrt{\frac{\rho}{\rho+2M}},0,0\right),\quad M^a=\left(\sqrt{\frac{\rho+2M}{\rho}},0,0,0\right),\nonumber\\
K&=&- \frac{M+2\rho}{\sqrt{\rho} (2M
+\rho)^{3/2}}. \label{e3A6}
\end{eqnarray}
The charge at the near-horizon limit $\rho \rightarrow 0$ is
\begin{eqnarray}
Q[\xi] = \frac{1}{8\pi G} \int_{\cal H} \sqrt{h} d^2x \left[\kappa T -\frac{1}{2} \partial_t T\right],\label{e3A7}
\end{eqnarray}
where $\kappa = \frac{1}{4M}= \frac{f'(r_h)}{2}$
is the surface gravity of the black hole. For
$T=T_m,T_n,$ the commutator [Eq.~(\ref{e27})] can
be directly calculated, and the result in the near-horizon limit $\rho \rightarrow 0$ is
\begin{eqnarray}
[Q_m,Q_n] &=& \frac{1}{8\pi G M} \int_{\cal H} \sqrt{h} d^2x \bigg[\kappa(T_m\partial_t T_n - T_n \partial_t T_m). \nonumber\\
&&\qquad\qquad\qquad -\frac{1}{2} (T_m \partial_t^2 T_n - T_n \partial_t^2 T_m)\nonumber\\
&&\qquad\qquad\qquad +\frac{1}{4\kappa}(\partial_t T_m \partial_t^2 T_n-\partial_t T_n \partial_t^2 T_m)\bigg].\label{e3A8}
\end{eqnarray}
In doing these calculations, we actually do not need
the explicit form of $f(r)$. From Eqs.~(\ref{e3A7}) and (\ref{e3A8}), we can see that
$f(\rho+r_h)$  only affects the surface gravity
$\kappa$; the second- and higher-order terms in
the expansion of $f(\rho+r_h)$ give no
contributions to the final results. So if we
start from the Rindler approximation of the full
metric, we can still get the same result.
We will show that this is also true for the
Gauss-Bonnet and third-order Lovelock cases
discussed in the next two subsections. The
extension to cases with a cosmological constant and
charges is straightforward. The charge and
Virasoro algebra will take the same forms as
Eqs.~(\ref{e3A7}) and (\ref{e3A8}), only with a
different surface gravity.

By substituting Eq.~(\ref{e310}) into
Eqs.~(\ref{e3A7}) and (\ref{e3A8}) and finishing the
integration over a cross-section area $A$, we
arrive at the final explicit expressions:
\begin{eqnarray}
&&Q_m =  \frac{A}{8\pi G} \frac{\kappa}{\alpha} \delta_{m,0},\nonumber\\
&&[Q_m, Q_n] = \frac{i \kappa A}{8\pi G \alpha} (m-n) \delta_{m+n,0} - i m^3 \frac{\alpha A}{16 \pi G \kappa} \delta_{m+n,0}. \label{e3A9}
\end{eqnarray}
Then we can obtain the central term in the algebra,
\begin{eqnarray}
K[\xi_m, \xi_n] &=&[Q_m, Q_n] + i (m-n) Q[\xi_{m+n}] \nonumber\\
&=& - i m^3 \frac{A}{16\pi G} \frac{\alpha}{\kappa} \delta_{m+n,0}.\label{e3A10}
\end{eqnarray}
From the central term, we can read off the central charge $C$ and the zero mode $Q_0$ as
\begin{eqnarray}
\frac{C}{12} = \frac{A}{16\pi G } \frac{\alpha}{\kappa},\qquad Q_0 \equiv Q[\xi_0] = \frac{A}{8\pi G} \frac{\kappa}{\alpha}.\label{e3A11}
\end{eqnarray}
Using the Cardy formula~\cite{Cardy:1986a,Bloete:1986a,Carlip:1998a}, we finally obtain the entropy
\begin{eqnarray}
S = 2\pi \sqrt{\frac{C Q_0}{6}} = \frac{A}{4G},\label{e3A12}
\end{eqnarray}
which is exactly the Bekenstein-Hawking entropy,
as expected. If we include a cosmological constant
or charge, the entropy formula will remain the
same.

\subsection{Entropy of a Gauss-Bonnet black hole}

Now we extend the approach to derive the entropy
of a Gauss-Bonnet black hole. The surface term
of the gravitation action now takes the
form~\cite{Myers:1987a}\cite{Davis:2003a}
\begin{eqnarray}
{\cal L}_B = K + 2\alpha_2 \left(J-2\hat{G}^{(1)}_{ab} K^{ab}\right),\label{e3B0}
\end{eqnarray}
where $\alpha_2$ is the coefficient of the
Gauss-Bonnet term, $\hat{G}^{(1)}_{ab}$ denotes
the $(n-1)$-dimensional Einstein tensor of the
induced metric $\gamma_{ab}$, and $J$ is the trace
of the following tensor
\begin{eqnarray}
J_{ab} = \frac{1}{3} \left(2 K K_{ac} K^c_{~b}+K_{cd} K^{cd} K_{ab} - 2 K_{ac} K^{cd} K_{db} -K^2 K_{ab}\right).\label{e3B1}
\end{eqnarray}
A family of general static spherical Gauss-Bonnet
black hole
solutions~\cite{Boulware:1985a}\cite{Cai:2001a}
takes the same form as Eq.~(\ref{e30}), but with
\begin{eqnarray}
f(r) = k +\frac{r^2}{4\alpha_2} \left(1-\sqrt{1+\frac{4}{3} \alpha_2 \Lambda + 8\alpha_2 \frac{\mu}{r^4}}\right).\label{e3B2}
\end{eqnarray}
The term $\Omega_{ij}$ denotes the metric of an $(n-2)$-dimensional hypersurface with the constant curvature
scalar $(n-2) (n-3) k$, $(k=0,\pm 1)$, and $\mu$ and
$\Lambda$ relate to the gravitational mass and the
cosmological constant, respectively. Without loss
of generality, we will focus on  the case with
$n=5$.

Now we proceed to calculate the charge and
Virasoro algebra for Gauss-Bonnet black holes
with different topologies. In the near-horizon
limit,  the charge and Virasoro algebra  can be
written in a unified form:
\begin{eqnarray}
&&Q[\xi] = \frac{1}{8\pi G} \left(1+\frac{12 k \alpha_2}{r_h^2}\right) \int_{\cal H} \sqrt{h} d^2x \left[\kappa T -\frac{1}{2} \partial_t T\right],\label{e3B3}\\
&&[Q_m,Q_n] = \frac{1}{8\pi G} \left(1+\frac{12 k \alpha_2}{r_h^2}\right) \int_{\cal H} \sqrt{h} d^2x \bigg[\kappa(T_m\partial_t T_n - T_n \partial_t T_m). \nonumber\\
&&\qquad\qquad\qquad\qquad\qquad\qquad -\frac{1}{2} (T_m \partial_t^2 T_n - T_n \partial_t^2 T_m)\nonumber\\
&&\qquad\qquad\qquad\qquad\qquad\qquad \left.+\frac{1}{4\kappa}(\partial_t T_m \partial_t^2 T_n-\partial_t T_n \partial_t^2 T_m)\right],\label{e3B4}
\end{eqnarray}
where the surface gravity
$\kappa=\frac{f'(r_h)}{2}$. During our
calculations, we keep all expansion terms in the
metric function  and  take the near-horizon limit
at the end. Comparing this with the Einstein case
[Eq.~(\ref{e3A8})], we see that an additional
factor $\left(1+\frac{12 k
\alpha_2}{r_h^2}\right)$ appears due to the
contribution of the Gauss-Bonnet term.

Choosing  $T_m$ as Eq.~(\ref{e310}) and substituting
it into the above equations, we get
\begin{eqnarray}
&&Q_m =  \frac{A}{8\pi G} \left(1+\frac{12 \alpha_2 k}{r_h^2}\right) \frac{\kappa}{\alpha} \delta_{m,0},\nonumber\\
&&[Q_m, Q_n] = - \frac{i \kappa A}{8\pi G  \alpha} \left(1+\frac{12 \alpha_2 k}{r_h^2}\right) (m-n) \delta_{m+n,0} - i m^3 \frac{\alpha A}{16 \pi G \kappa} \left(1+\frac{12\alpha_2 k }{r_h^2}\right)\delta_{m+n,0}.\label{e3B5}
\end{eqnarray}
The central term reads
\begin{eqnarray}
K[\xi_m, \xi_n] &=&[Q_m, Q_n] + i (m-n) Q[\xi_{m+n}] \nonumber\\
&=& - i m^3 \frac{A}{16\pi
G}\left(1+\frac{12\alpha_2 k }{r_h^2}\right)
\frac{\alpha}{\kappa} \delta_{m+n,0},\label{e3B6}
\end{eqnarray}
from which we can read off the central charge $C$
and zero mode $Q_0$ as
\begin{eqnarray}
\frac{C}{12} =\frac{A}{16\pi G} \left(1+\frac{12 \alpha_2 k}{r_h^2}\right)\frac{\alpha}{\kappa},\qquad Q_0 \equiv Q[\xi_0] = \frac{A}{8\pi G} \left(1+\frac{12 \alpha_2 k}{r_h^2}\right)\frac{\kappa}{\alpha}.\label{e3B7}
\end{eqnarray}
Using the Cardy formula, we obtain the entropy
\begin{eqnarray}
S = 2\pi \sqrt{\frac{C Q_0}{6}} = \frac{A}{4G} \left(1+\frac{12 \alpha_2 k}{r_h^2}\right),\label{e3B8}
\end{eqnarray}
which is exactly the Wald entropy of the Gauss-Bonnet black
holes~\cite{Wald:1993a,Iyer:1994a,Iyer:1995a,Jacobson:1993a,Clunan:2004a}.
For the flat case with $k=0$, the entropy in Eq.~(\ref{e3B8}) reduces to
the Bekenstein-Hawking entropy as expected, because at this time
$\Omega_{ij} dx^i dx^j$ is flat and will not cause corrections to
the Bekenstein-Hawking entropy. This result coincides with the
discovery in Ref.~\cite{Clunan:2004a} from the Wald entropy formula or
thermodynamics approach.

\subsection{Entropy of a third-order Lovelock black hole}

Let us go further to see if this approach can
produce the correct entropy for a more general
gravitational theory, the third-order Lovelock
gravity.  The boundary term now reads
as~\cite{Dehghani:2006a}
\begin{eqnarray}
{\cal L}_B &=& K + 2\alpha_2 \left(J-2\hat{G}^{(1)}_{ab} K^{ab}\right)
+ 3 \alpha_3 \left(P-2\hat{G}^{(2)}_{ab} K^{ab} - 12 \hat{R}_{ab} J^{ab}\right.\nonumber\\&&
\left.+2 \hat{R} J -4 K \hat{R}_{abcd} K^{ac} K^{bd} - 8 \hat{R}_{abcd} K^{ac} K^b_{~e} K^{ed}\right).\label{e3C0}
\end{eqnarray}
Here, $\hat{R}_{abcd}$ is the intrinsic Riemannian
tensor constructed from the boundary metric
$\gamma_{ab}$. $\hat{G}^{(2)}_{ab}$ is the second-order Lovelock tensor of $\gamma_{ab}$; that is,
\begin{eqnarray}
\hat{G}^{(2)}_{ab} = 2\left(\hat{R}_{acde}
\hat{R}_b^{~cde}-2 \hat{R}_{acbd} \hat{R}^{cd} -2
\hat{R}_{ac} \hat{R}^c_{~b} +\hat{R}
\hat{R}_{ab}\right) -\frac{1}{2} {\cal L}_2
\gamma_{ab},\label{e3C1}
\end{eqnarray}
with ${\cal L}_2$ being the Gauss-Bonnet term ${\cal
L}_2 = \hat{R}_{abcd}
\hat{R}^{abcd}-4\hat{R}_{ab} \hat{R}^{ab}
+\hat{R}^2$. $P$ is the trace of the following
tensor:
\begin{eqnarray}
P_{ab}&=&\frac{1}{5} \bigg\{\left[K^4-6 K^2 K^{cd} K_{cd} + 8 K K_{cd} K^d_{~e} K^{ec} -6 K_{cd} K^{de} K_{ef} K^{fc} +3\left(K_{cd} K^{cd}\right)^2\right] K_{ab}\nonumber\\
&&\qquad-\left(4K^3-12K K_{ed} K^{ed} +8 K_{de} K^e_{~f} K^{fd}\right)K_{ac} K^c_{~b} -24 K K_{ac} K^{cd} K_{de} K^e_{~b}\nonumber\\
&&\qquad +\left(12K^2-12 K_{ef} K^{ef}\right)K_{ac} K^{cd} K_{db} +24 K_{ac} K^{cd} K_{de} K^{ef} K_{bf}\bigg\}.\label{e3C2}
\end{eqnarray}
Without loss of generality, we will concentrate
on the seven-dimensional static, spherical black
hole solutions of the third-order Lovelock
gravity. General black hole solutions in third-order Lovelock gravity with arbitrary
coefficients $\alpha_2$ and $\alpha_3$ are rather
complicated to obtain. In Refs.~\cite{Dehghani:2005a,Dehghani:2009a}, it was found
that an explicit solution of a black hole exists
when the two coefficients satisfy $2
\alpha_2^2=\alpha_3=\frac{\tilde{\alpha}^2}{72}$
with the metric function in Eq.~(\ref{e30}) in
the form
\begin{eqnarray}
f(r) = k + \frac{r^2}{\tilde{\alpha}}
\left[1-\left(1+\frac{\Lambda
\tilde{\alpha}}{5}+\frac{3\tilde{\alpha} \mu}{5
r^6}\right)^{1/3}\right].\label{e3C3}
\end{eqnarray}
The term $\Omega_{ij} dx^i dx^j$ denotes an $(n-2)$-dimensional hypersurface with the constant curvature
scalar $(n-2) (n-3) k$, $(k=0,\pm 1)$. Following the
approach we discussed above,  we can obtain
unified expressions of the charge and the
Virasoro algebra for different topological
spacetimes:
\begin{eqnarray}
&&Q[\xi] = \frac{1}{8\pi G} \left(1+\frac{40 \alpha_2 k}{r_h^2}+\frac{360 \alpha_3 k^2}{r_h^4}\right) \int_{\cal H} \sqrt{h} d^2x \left[\kappa T -\frac{1}{2} \partial_t T\right],\label{e3C4}\\
&&[Q_m,Q_n] = \frac{1}{8\pi G} \left(1+\frac{40\alpha_2 k }{r_h^2}+\frac{360 \alpha_3 k^2}{r_h^4}\right) \int_{\cal H} \sqrt{h} d^2x \bigg[\kappa(T_m\partial_t T_n - T_n \partial_t T_m). \nonumber\\
&&\qquad\qquad\qquad\qquad\qquad\qquad \qquad \qquad\qquad  -\frac{1}{2} (T_m \partial_t^2 T_n - T_n \partial_t^2 T_m)\nonumber\\
&&\qquad\qquad\qquad\qquad\qquad\qquad\qquad \qquad\qquad   \left.+\frac{1}{4\kappa}(\partial_t T_m \partial_t^2 T_n-\partial_t T_n \partial_t^2 T_m)\right],\label{e3C5}
\end{eqnarray}
where the surface gravity
$\kappa=\frac{f'(r_h)}{2}$, and the additional
factor changes to be $\left(1+\frac{40 \alpha_2
k}{r_h^2}+\frac{360 \alpha_3 k^2}{r_h^4}\right)$,
containing the contribution of the third Lovelock
term. After using the Cardy formula, we obtain
the entropy of the third Lovelock black hole,
\begin{eqnarray}
S = \frac{A}{4G} \left(1+\frac{40\alpha_2 k }{r_h^2}+\frac{360 \alpha_3 k^2}{r_h^4}\right),\label{e3C6}
\end{eqnarray}
which is consistent with the result derived in Ref.~\cite{Dehghani:2005a,Dehghani:2009a} in the form
of the Wald entropy formula. Here we cannot
expect the above entropy to reduce to the
Gauss-Bonnet case [Eq.~(\ref{e3B8})] by taking the
limit $\alpha_3=0$, because in our considered case,
$\alpha_2$ and $\alpha_3$ are not independent.

Although we only consider a special case with two
coefficients satisfying a particular relation, we
want to indicate that in calculating the Noether
charge [Eq.~(\ref{e3C4})] and Virasoro algebra
[Eq.~\ref{e3C5})], we actually do not need the explicit
form of the function $f(r)$ in Eq.~(\ref{e3C3})-what we need is only the form of the metric
Eq.~(\ref{e30}). So, for more general solutions of
the black hole with the metric in the form of
Eq.~(\ref{e30}), we can get a similar result
for different surface gravity and different
correction terms in the entropy expression.

\section{summary}

In this paper, we have generalized the method
proposed in Refs.~\cite{Majhi:2012a,Majhi:2012b} to
derive the entropy of black holes in Gauss-Bonnet
and third-order Lovelock gravity. We have shown
that in addition to reproducing the correct
Bekenstein-Hawking entropy in the Einstein
gravity, the method can produce the Wald entropy
in the general theories of gravity with higher-curvature corrections. This shows that the
approach based on the Virasoro algebra and
central charge from the surface term of
gravitational action is general, which can
provide deeper insight into the nature of the
entropy.

In our calculation we noticed that  the
expressions of Noether charge and Virasoro
algebra only depend on the form of the metric
[Eq.~(\ref{e30})], and do not heavily depend on the
explicit expression in the metric function. This
tells us that the method to derive the entropy
can even work  for more complicated expressions
in the metric form of Eq.~(\ref{e30}). It is interesting
to ask whether the method is still effective for
rotating black holes. More investigation in
this direction is called for, as it can show us
the universality of the method.

\end{document}